\documentclass[a4paper,11pt]{article}
\usepackage{pos}
\usepackage{physics}

\title{Calculation of the pion charge radius from an improved model-independent method}

\author*[a]{Kohei Sato}
\author[b]{Hiromasa~Watanabe}
\author[c,d]{Takeshi~Yamazaki}

\affiliation[a]{Degree Programs in Pure and Applied Sciences, Graduate School of Science and Technology, University of Tsukuba, Tsukuba, Ibaraki 305-8571, Japan}
\affiliation[b]{Yukawa Institute for Theoretical Physics, Kyoto University, Kyoto 606-8502, Japan}
\affiliation[c]{Faculty of Pure and Applied Sciences, University of Tsukuba, Tsukuba, Ibaraki 305-8571, Japan}
\affiliation[d]{Center for Computational Sciences, University of Tsukuba, Tsukuba, Ibaraki 305-8577, Japan}

\emailAdd{ksatoh@het.ph.tsukuba.ac.jp}

\abstract{
We propose a new improved model-independent method for calculating the pion charge radius.
In a recently-proposed model-independent method for the pion charge radius,
we find it difficult to compute the pion charge radius for small pole mass $M_{\rm{pole}}^2$ and volume
due to systematic errors coming from finite volume effect and higher-order contamination of the Taylor expansion of the form factor.
We circumvent this difficulty by introducing a new appropriate function and
propose a modified method that can calculate the pion charge radius with less systematic errors
in the small $M_{\rm{pole}}^2$ and volume cases.
As preliminary results, we check that our improved model-independent method works well
on a mockup data and also an actual lattice QCD data at the pion mass of 0.51 GeV.
}

\FullConference{%
The 39th International Symposium on Lattice Field Theory,\\
8th-13th August, 2022,\\
Rheinische Friedrich-Wilhelms-Universit\"{a}t Bonn, Bonn, Germany
}

\begin{document}
\maketitle

\section{Introduction}

In particle physics, quark is considered to be a charged point particle with no spatial extent,
whereas hadron is a bound state of quarks and gluons, which are held together by strong interaction,
and has an internal structure with spatial extent.
One of the intrinsic properties of the hadron is the charge radius,
which corresponds to the spatial extent of hadron's charge distribution.

\par

The simplest hadron, the pion, is composed of up and down quarks (and their respective antiparticles).
The charge radius of the pion $\expval{r_{\pi}^2}$ is defined in relation to the form factor of the pion electromagnetic vertex.
The pion form factor $F_\pi(q^2)$ is given by the matrix element
\begin{eqnarray}
	\matrixel{\pi^{+}(p_{f})}{V_{\mu}}{\pi^{+}(p_{i})}=(p_{f}+p_{i})_{\mu}F_{\pi}(q^2) ,
\end{eqnarray}
where $q^2=-(p_{f}-p_{i})^2\geq0$ is the momentum transfer,
$V_{\mu}=\sum_{f}Q_{f}\bar{\psi}_{f}\gamma_{\mu}\psi_{f}$ is the electro-magnetic vector current
with the flavor index $f$ and $Q_{f}$ is the charge of the $f$ quark.
The pion charge radius is defined as
\begin{eqnarray}
	\expval{r_{\pi}^2}=-6\eval{\dv{q^2}F_{\pi}(q^2)}_{q^2=0}
	\label{eq:pion_char_radi} ,
\end{eqnarray}
using the first derivative of the pion form factor.
The pion charge radius has long been obtained experimentally,
and nowadays it is measured precisely with an experimental error of about $1\%$~\cite{ParticleDataGroup:2022pth}.
On the other hand, it has also been calculated from lattice QCD.
Although the charge radius has been measured with an error of about $3\%$ in lattice QCD,
their results have large error compared to experiment,
or do not agree with experimental value within the error~\cite{Gao:2021xsm}.
The accurate determination of the charge radius from lattice QCD calculation gives us useful information
about not only the size and structure of the hadron but also crucial precision tests of the Standard Model at low energy.
From this point of view, it is very important to study a crucial method of  calculating the pion charge radius in lattice QCD.

\par

One of the reasons for the large lattice QCD error is that
the traditional method of calculating the pion charge radius uses some fit ansatzes.
The fit ansatz gives systematic error and increases the lattice QCD error.
Therefore, a model-independent method that does not use the fit ansatz was researched~\cite{Aglietti:1994nx,Lellouch:1994zu,Bouchard:2016gmc}.
Recently, a paper has been published that improves this model-independent method and applies it to the calculation of the pion charge radius~\cite{Feng:2019geu}.
The purpose of this work is investigating properties of the model-independent method and improving the method further.

\section{Traditional method}
We briefly explain the traditional method for obtaining the charge radius.
First, we make the form factor data for each momentum transfer $q^2$ from the equation,
\begin{eqnarray}
	F_{\pi}(q^2)=\dfrac{2E_{\pi}(\vec{p})Z_{\pi}(\vec{0})}{(E_{\pi}(\vec{p})+m_{\pi})Z_{\pi}(\vec{p})}
					\dfrac{\tilde{C}_{\pi V\pi}(t,t_{\rm{sink}};\vec{p})}{
							\tilde{C}_{\pi V\pi}(t,t_{\rm{sink}};\vec{0})}e^{(E_{\pi}(\vec{p})-m_{\pi})t} ,
\end{eqnarray}
using the 3-point function in $0 \ll t \ll t_{\rm sink}$
\begin{eqnarray}
	\tilde{C}_{\pi V\pi}(t,t_{\rm{sink}};\vec{p})=Z_{V}\sum_{\vec{x},\vec{y},\vec{z}}
	\matrixel{0}{\pi^{+}(\vec{z},t_{\rm{sink}})V_{4}(\vec{y},t){\pi^{+}}^{\dag}(\vec{x},0)}{0}e^{i\vec{p}\cdot(\vec{x}-\vec{y})},
\end{eqnarray}
where $\vec{p}:=\frac{2\pi}{L}(n_{x}, n_{y}, n_{z})$ with $n_i$ and $L$ being an integer and the spatial extent,
$E_{\pi}(\vec{p}):=\sqrt{m_{\pi}^2+\vec{p}^2}$,
$Z_{\pi}(\vec{p}):=\matrixel{0}{\pi^{+}(\vec{0},0)}{E_{\pi}(\vec{p})}$
and $Z_{V}$ is a renormalization constant of the vector current.
Second, we fit the form factor data using a fit ansatz, such as the monopole formula or quadratic formula given by,
\begin{eqnarray}
	F_{\pi}(q^2)=\dfrac{1}{1+q^2/M_{\rm{pole}}^2} , \hspace{20pt}
	F_{\pi}(q^2)=1+f_{1}q^2+f_{2}(q^2)^2 \label{eq:fit ansatz} ,
\end{eqnarray}
where $M_{\rm{pole}}^2$ and $f_{1}, f_{2}$ are parameters.
Third, we differentiate the form factor using the fit result and Eq.~(\ref{eq:pion_char_radi}).
From the above procedure, the pion charge radius is obtained.

\par

In this method, it is assumed that the form factor obeys Eq.~(\ref{eq:fit ansatz}) .
This hypothesis is a model-dependent part and one of the sources for systematic errors in the calculation.

\section{Model-independent method}
To reduce the systematic error in the traditional method,
model-independent method was proposed~\cite{Aglietti:1994nx,Lellouch:1994zu,Bouchard:2016gmc}
and its improvement for the pion charge radius was investigated~\cite{Feng:2019geu}.
In this section, we briefly review the method proposed in these papers
and show how to calculate the pion charge radius using the model-independent method.

\subsection{Model-independent method for the continuum limit and infinite volume}
In the continuum limit and infinite volume ($a\to0,\,V\to\infty$),
\begin{eqnarray}
	\dv[n]{\tilde{F}(\vec{p})}{{(\abs{\vec{p}}^{2})}}
		=(-1)^{n}\dfrac{n!}{(2n+1)!}\int \dd[3]{x}\abs{\vec{x}}^{2n}F(\vec{x})
		\label{eq:3dimmoment}
\end{eqnarray}
is obtained for a function $F(\vec{x})$ satisfying $F(\vec{x})=F(\abs{\vec{x}})$,
where $\tilde{F}(\vec{p})$ is the Fourier transform of $F(\vec{x})$.
The meaning of this equation is that
the $n$th-order $|\vec{p}|^2$ derivative at $\abs{\vec{p}}^{2}=0$ is equal to the $2n$th-order spatial moment.
In other words, calculating the pion charge radius can be translated into calculating the $2n$th-order spatial moment.

\par

To show the model independence using Eq.~(\ref{eq:3dimmoment}),
we consider a relationship between a 1-dimension 3-point function and the charge radius as an example.
The 1-dimension 3-point function is defined by
\begin{eqnarray}
	C_{\pi V\pi}(t,t_{\rm{sink}};r):=Z_{V}\sum_{\vec{z}}\sum_{y_{2},y_{3}}\sum_{x_{2},x_{3}}
	\matrixel{0}{\pi^{+}(\vec{z},t_{\rm{sink}})V_{4}(\vec{y},t){\pi^{+}}^{\dag}(\vec{x},0)}{0}
	\label{eq:1pt3func} ,
\end{eqnarray}
where $r:=\abs{x_{1}-y_{1}}$.
Periodic boundary condition is imposed in all spacetime directions.
The spatial Fourier transform of Eq.~(\ref{eq:1pt3func}) in $0\ll t \ll t_{\rm{sink}}$
normalized by that with $p=0$ yields
the momentum 3-point function of the ground state as,
\begin{eqnarray}
	\tilde{C}_{\pi V\pi}(t;p)
	=\dfrac{\tilde{C}_{\pi V\pi}(t,t_{\rm{sink}};p)}{\tilde{C}_{\pi V\pi}(t,t_{\rm{sink}};0)}
	=\dfrac{Z_{\pi}(p)}{Z_{\pi}(0)}\dfrac{E_{\pi}(p)+m_{\pi}}{2E_{\pi}(p)}
		F_{\pi}(q^2)e^{-(E_{\pi}(p)-m_{\pi})t} .
\end{eqnarray}
The momentum derivative of this function is
\begin{eqnarray}
	\eval{\dv{\tilde{C}_{\pi V\pi}(t;p)}{p^2}}_{p^2=0}
	=\qty(\eval{\dv{q^2}F_{\pi }(q^2)}_{q^2=0} {\mbox{term}})+\qty({\mbox{known factors}})
	\label{eq:dvCpiVpi(p)} ,
\end{eqnarray}
where the known factors represent the differential terms other than the form factor.
Recalling that the first-order derivative of the form factor is equal to the charge radius and
that the momentum derivative is equal to the spatial moment,
we can see that the charge radius is obtained by only calculating the $r^2$ moment of the 3-point function.

\par

In this example, we have not assumed the fit ansatz,
and hence,
this method is model independent.
Note, however, that Eq.~(\ref{eq:dvCpiVpi(p)}) holds in $a\to0$ and $V\to\infty$.

\subsection{Model-independent method for finite volume} \label{subsec:Model-inde.finite volume}
We consider the model-independent method on a finite volume.
For finite volume, higher-order contamination arises from finite volume effect in this method as explained below.
To see this finite volume effect, we consider the $r^{2n}$ moment of the 3-point function:
\begin{eqnarray}
	C^{(n)}(t):=\dfrac{\sum_{r}r^{2n}C_{\pi V\pi}(t,t_{\rm{sink}};r)}{\tilde{C}_{\pi V\pi}(t,t_{\rm{sink}};0)}.
	\label{eq:1ptr^2n3func}
\end{eqnarray}
The spatial Fourier transform of the 3-point function $C_{\pi V\pi}(t,t_{\rm{sink}};r)$ yields
\begin{eqnarray}
	C^{(n)}(t)=\sum_{p}\Delta(t,p)T_{n}(p)F_{\pi}(q^2)
	\label{eq:sumpDTF} ,
\end{eqnarray}
where
\begin{eqnarray}
	\Delta(t,p):=\dfrac{Z_{\pi}(p)}{Z_{\pi}(0)}\dfrac{E_{\pi}(p)+m_{\pi}}{2E_{\pi}(p)}
							e^{-(E_{\pi}(p)-m_{\pi})t} , \hspace{10pt}
	T_{n}(p):=\dfrac{1}{L}\sum_{r}r^{2n}e^{ipr} .
\end{eqnarray}
From the Taylor expansion $F_{\pi}(q^2)=\sum_{m=0}^{\infty}f_{m}q^{2m}$ of the form factor,
Eq.~(\ref{eq:sumpDTF}) can be written as
\begin{eqnarray}
	\hspace{-50pt}
	C^{(n)}(t)=f_{0}\beta_{0,n}(t)+f_{1}\beta_{1,n}(t)+
		f_{2}\beta_{2,n}(t)+\cdots=\sum_{m=0}^{\infty}f_{m}\beta_{m,n}(t)
	\label{eq:cn=fbeta} ,
\end{eqnarray}
where the function $\beta_{m,n}(t)$ is a known function given by,
\begin{eqnarray}
	\beta_{m,n}(t):=\sum_{p}\Delta(t,p)T_{n}(p)q^{2m} .
\end{eqnarray}

\par

From Eqs.~(\ref{eq:3dimmoment}) and (\ref{eq:dvCpiVpi(p)}), in the case of infinite volume,
only the first-order derivative of the form factor is obtained from $C^{(1)}(t)$.
On the other hand, from Eq.~(\ref{eq:cn=fbeta}), in the case of finite volume,
not only the first-order derivative $f_{1}$ but also the terms of the higher-order derivatives,
such as $f_{2}$ and $f_{3}$, remain in $C^{(1)}(t)$.
These terms are caused by the finite volume effect, which leads to systematic errors in the calculation of the charge radius.

\par

To reduce the higher-order contamination, the function $R(t)$ is defined by
\begin{eqnarray}
	R(t)&:=&\alpha_{1}C^{(1)}(t)+\alpha_{2}C^{(2)}(t)+h\notag\\
	&=&(\alpha_{1}\beta_{0,1}+\alpha_{2}\beta_{0,2}+h)
		+(\alpha_{1}\beta_{1,1}+\alpha_{2}\beta_{1,2})f_{1}
		+(\alpha_{1}\beta_{2,1}+\alpha_{2}\beta_{2,2})f_{2}+\cdots
	\label{eq:defR(t)} ,
\end{eqnarray}
where we use $f_{0}=1$ and the dots represent higher-order terms with $f_m$ ($m\ge 3$).
The parameters $\alpha_{1},\alpha_{2},h$ are defined to satisfy
\begin{eqnarray}
	\alpha_{1}\beta_{0,1}+\alpha_{2}\beta_{0,2}+h=0,
	\hspace{15pt}\alpha_{1}\beta_{1,1}+\alpha_{2}\beta_{1,2}=1,
	\hspace{15pt}\alpha_{1}\beta_{2,1}+\alpha_{2}\beta_{2,2}=0
	\label{eq:defParaAlphaH} .
\end{eqnarray}
From Eqs.~(\ref{eq:defR(t)}) and (\ref{eq:defParaAlphaH}), the function $R(t)$ is rewritten as,
\begin{eqnarray}
	R(t)=f_{1}+\sum_{m=3}^{\infty}\qty(\sum_{k=1}^{2}\alpha_{k}\beta_{m,k}(t))f_{m}.
	\label{eq:R(t)}
\end{eqnarray}
The first term is a constant term and the second term is a time-dependent term.
The second time-dependent term is the higher-order contamination term, which gives rise to the finite volume effect.
If the higher-order contamination term is small, we can obtain the charge radius from the constant term in $R(t)$.
In this paper, we call this method original model-independent method.

\subsection{Properties of the original method} \label{subsec:problem}
As explained in the previous subsection, the original model-independent method contains the higher-order contamination due to finite volume effect.
We discuss systematic errors coming from the effect using mockup data.
In particular, we consider analyses with the pion form factor expressed by the monopole formula, 
\begin{eqnarray}
	F_{\pi}(q^2)=\dfrac{1}{1+q^2/M_{\rm{pole}}^2} .
	\label{eq:monopole_form}
\end{eqnarray}
This is because, from vector meson dominance model, the pion form factor is well represented by the monopole formula~\cite{Gao:2021xsm}.
To see the systematic error of the contamination in the analysis of the pion 3-point function,
we make the following mockup data of $C^{(n)}(t)$ using
\begin{equation}
    C^{(n)}(t)=\sum_{p}\Delta(t,p)T_{n}(p)F(q^2),
\end{equation}
where
\begin{equation}
	m_{\pi}=0.25, \hspace{10pt} \Delta(t,p)=\dfrac{E(p)+m_{\pi}}{2E(p)}e^{-(E(p)-m_{\pi})t},\hspace{10pt}
	T_{n}(p)=\dfrac{1}{L}\sum_{x}x^{2n}\cos(px) .
\end{equation}

\par

\begin{figure}[!th]
 \centering
 \includegraphics[width=75mm,pagebox=cropbox]{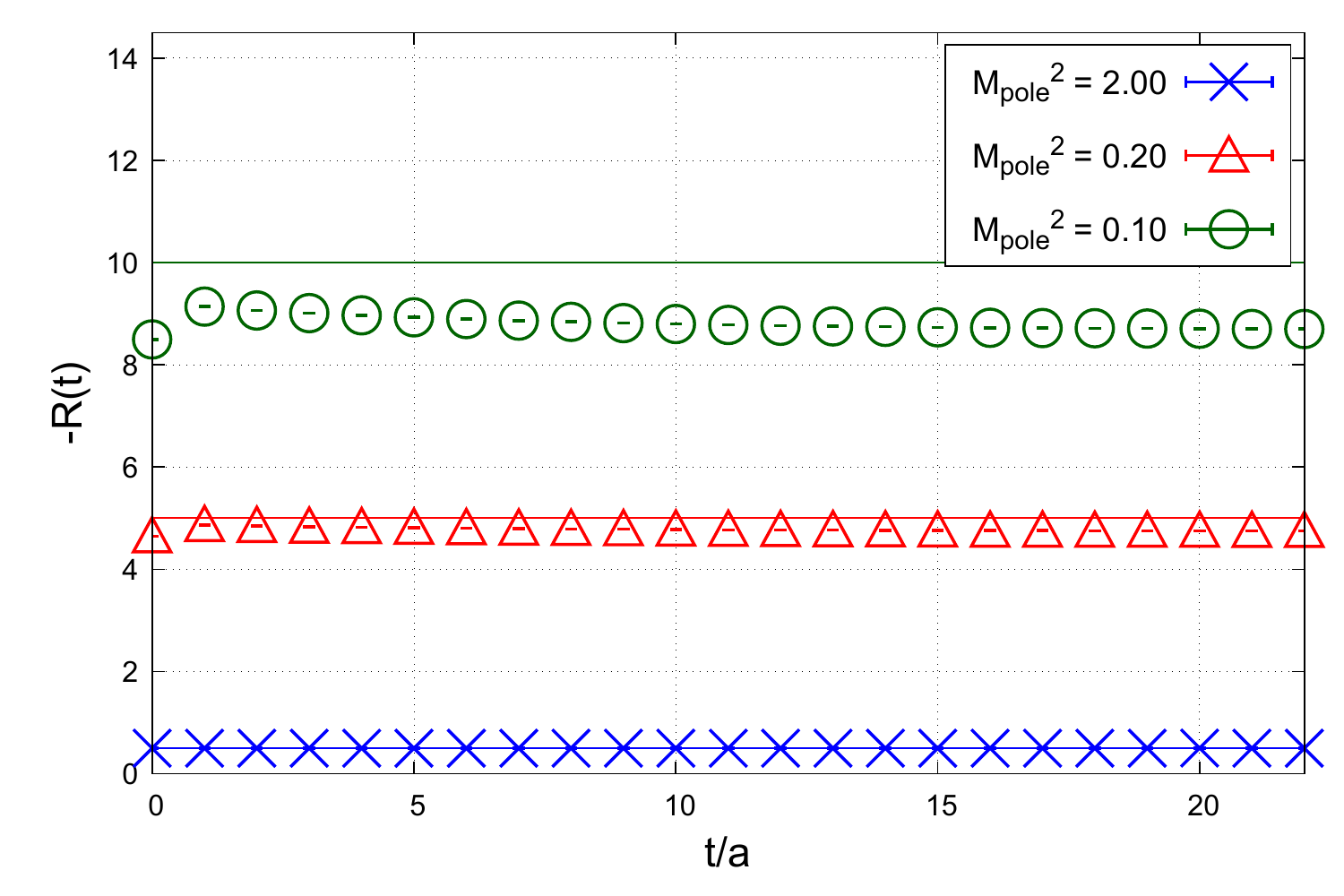}
 \includegraphics[width=75mm,pagebox=cropbox]{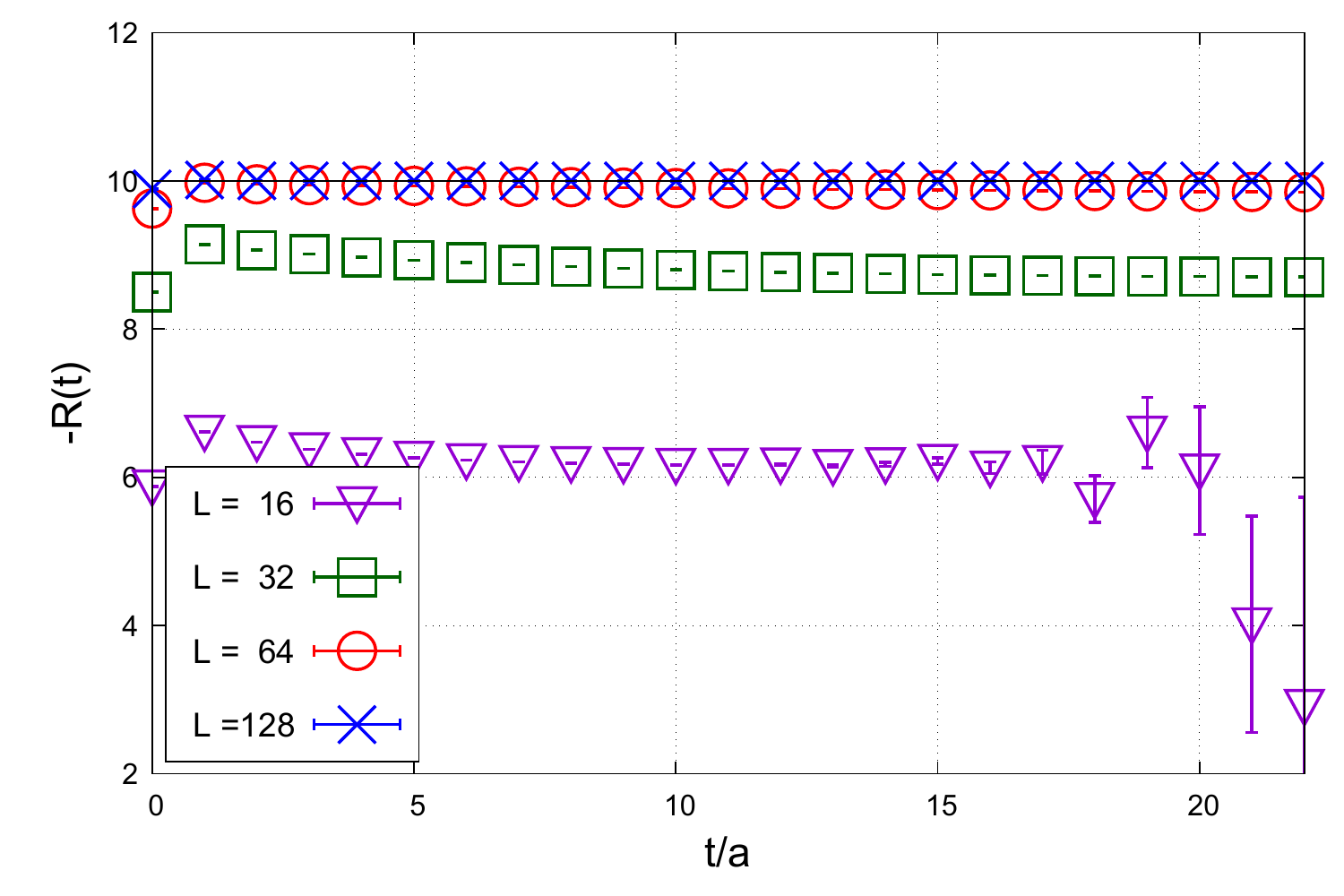}
 \caption{
Results of analyzing the mockup data for various $M_{\rm{pole}}^2$ and volume using the original method.
Symbols are calculation results, and solid lines are exact values of $f_{1}=-1/M_{\rm{pole}}^2$.
Left: $M_{\rm{pole}}^2$ dependence for $M_{\rm{pole}}^2 = 0.10, 0.20, 2.00$ with the fixed volume of $L=32$.
Right: Volume dependence for $L=16, 32, 64, 128$ with the fixed $M_{\rm pole}^2 = 0.10$.
  \label{fig:R(t)_Mpole_V_NotImp}
 }
\end{figure}

We analyze the mockup data for various $M_{\rm{pole}}^2$ and volumes using the original method of Eq.~(\ref{eq:defR(t)}).
The left panel of Fig.~\ref{fig:R(t)_Mpole_V_NotImp} shows the $M_{\rm{pole}}^2$ dependence of the original method.
The mockup data is analyzed in three cases
for $M_{\rm{pole}}^2 = 0.10,\hspace{2pt}0.20,\hspace{2pt}2.00$ with the fixed volume $L=32$.
For the largest $M_{\rm{pole}}^2$, the calculation result (symbol) is consistent with the exact value (solid line),
but, for small $M_{\rm{pole}}^2$, the result is not consistent.
Moreover, the right panel of Fig.~\ref{fig:R(t)_Mpole_V_NotImp} shows the volume dependence of the original method.
The mockup data is analyzed in four cases
for $L=16,\hspace{2pt}32,\hspace{2pt}64,\hspace{2pt}128$ with fixed $M_{\rm{pole}}^2=0.10$.
For large volume, the calculation result is consistent with the exact value,
but, for small volume, the result is not consistent.
These differences are caused by the higher-order contamination.
Therefore, it is hard for the original method to compute $f_{1}$ at small $M_{\rm{pole}}^2$ and volume.

\par

To see the cause of the problem, we consider the Taylor expansion of the monopole formula as,
\begin{equation}
	F_{\pi}(q^2) = \dfrac{1}{1+q^2/M_{\rm{pole}}^2}=\sum_{m=0}^{\infty}\qty(-\dfrac{1}{M_{\rm{pole}}^2})^{m}q^{2m} ,
\end{equation}
where the coefficient $f_{m}=(-1/M_{\rm{pole}}^2)^{m}$ in this case.
If $M_{\rm{pole}}^2<1$ then $\abs{f_{m}}$ increases as $m$.
From this property of the coefficient $f_{m}$ and the explicit form of the higher-order contamination in Eq.~(\ref{eq:R(t)}),
we can explain why the original method dose not work well for small $M_{\rm{pole}}^2$ and volume.
For small volume, the contamination of higher-order $\beta_{m,1},\hspace{2pt}\beta_{m,2}\hspace{3pt}(m\geq3)$
remains in $R(t)$ due to finite volume effect,
and for small $M_{\rm{pole}}^2$, the convergence of the expansion becomes poor, since $|f_{m}|$ increases as $m$.
Therefore, for small $M_{\rm{pole}}^2$ and volume,
the higher-order contamination in the second term of Eq.~(\ref{eq:R(t)}) causes sizable systematic errors.
This problem may also occur in actual lattice QCD data and the original method needs to be improved in such cases.

\subsection{A new improved model-independent method}
To reduce the higher-order contamination for small $M_{\rm{pole}}^2$ and volume,
we propose a new model-independent method.
Our basic idea is to improve the convergence of the coefficient $f_{m}$.
Our improved model-independent method introduces an appropriate function $G(q^2)$
to replace the bad convergent function $F_{\pi}(q^2)$ with the good one $S(q^2):=F_{\pi}(q^2)G(q^2)$.
Using this idea,
the higher-order contamination corresponding to the second term in Eq.~(\ref{eq:R(t)}) can be reduced more.

\par

In Eq.~(\ref{eq:sumpDTF}), inserting a function $G(q^2)$ yields
\begin{eqnarray}
	C^{(n)}(t)=\sum_{p}\Delta(t,p)T_{n}(p)S(q^2)\dfrac{1}{G(q^2)}
	\label{eq:sumpDTS/G} .
\end{eqnarray}
From the Taylor expansion of $S(q^2)$, $S(q^2)=\sum_{m=0}^{\infty}s_{m}q^{2m}$,
Eq.~(\ref{eq:sumpDTS/G}) can be written as
\begin{eqnarray}
	C^{(n)}(t)=s_{0}\tilde{\beta}_{0,n}(t)+s_{1}\tilde{\beta}_{1,n}(t)+
		s_{2}\tilde{\beta}_{2,n}(t)+\cdots=\sum_{m=0}^{\infty}s_{m}\tilde{\beta}_{m,n}(t)
	\label{eq:exp_Cn_new} ,
\end{eqnarray}
where $\tilde{\beta}_{m,n}(t)$ is a known function,
\begin{eqnarray}
	\tilde{\beta}_{m,n}(t):=\sum_{p}\Delta(t,p)T_{n}(p)q^{2m}/G(q^2).
\end{eqnarray}
To improve the convergence of $f_{m}$, we consider $G(q^2):=1+g_{1}q^2+g_{2}q^4$,
where parameters $g_{1},\hspace{2pt}g_{2}$ are chosen to satisfy
\begin{eqnarray}
	s_{2}=f_{2}+f_{1}g_{1}+f_{0}g_{2}=0 .
\end{eqnarray}
The reason is that if the form factor is well represented by the monopole formula (\ref{eq:monopole_form}),
then 
\begin{eqnarray}
	s_{m}&\sim&(-1/M_{\rm{pole}}^2)s_{m-1} \hspace{10pt}(m\geq3)
\end{eqnarray}
holds for the coefficients $s_{m}$ of the function $S(q^2)$,
and the convergence of the expansion (\ref{eq:exp_Cn_new}) becomes better due to $s_m \sim 0$ for $m \ge 3$.
Other procedures to determine the charge radius are the same as in Eqs.~(\ref{eq:defR(t)})-(\ref{eq:R(t)}).

\par

\begin{figure}[!th]
 \centering
 \includegraphics[width=75mm,pagebox=cropbox]{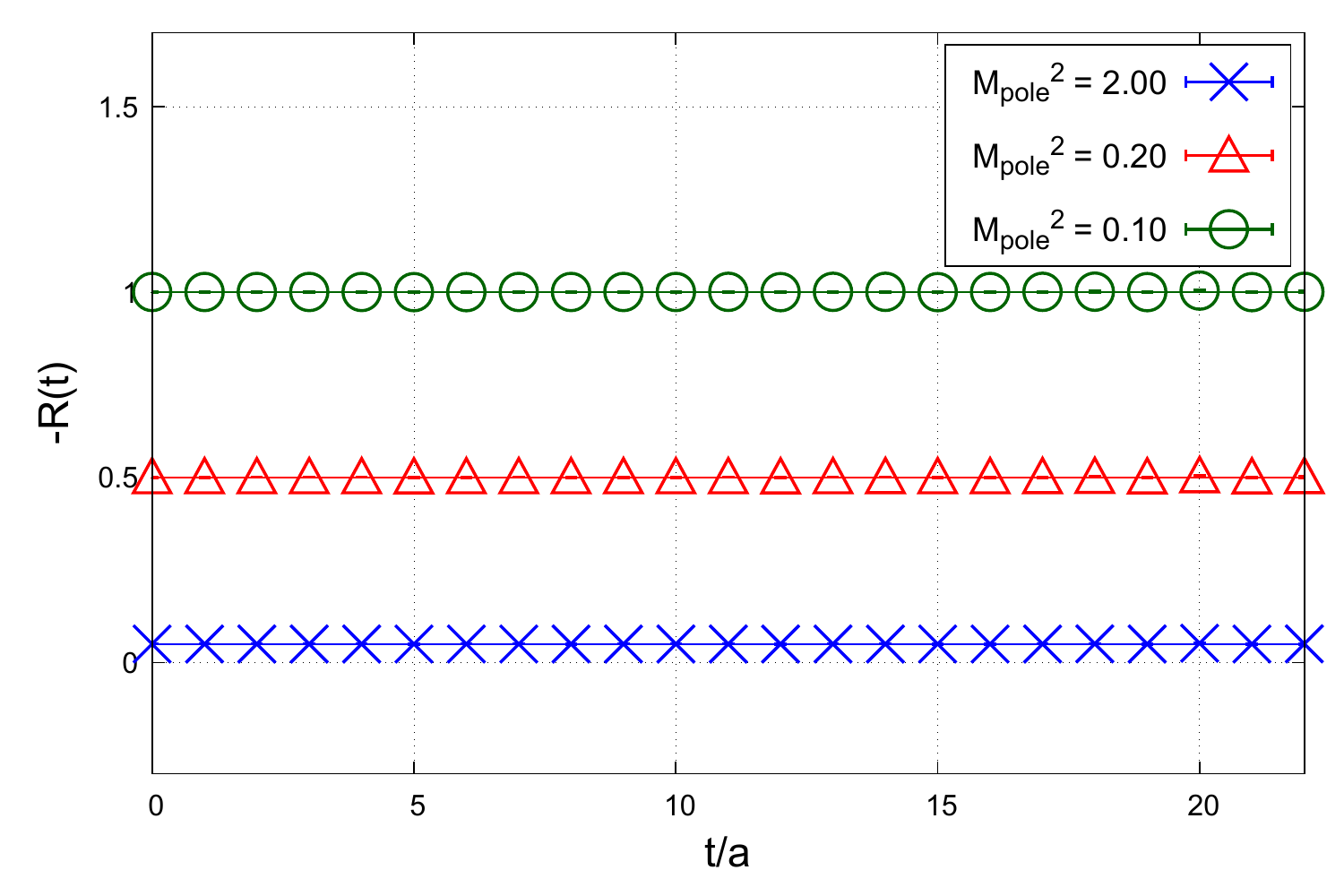}
 \includegraphics[width=75mm,pagebox=cropbox]{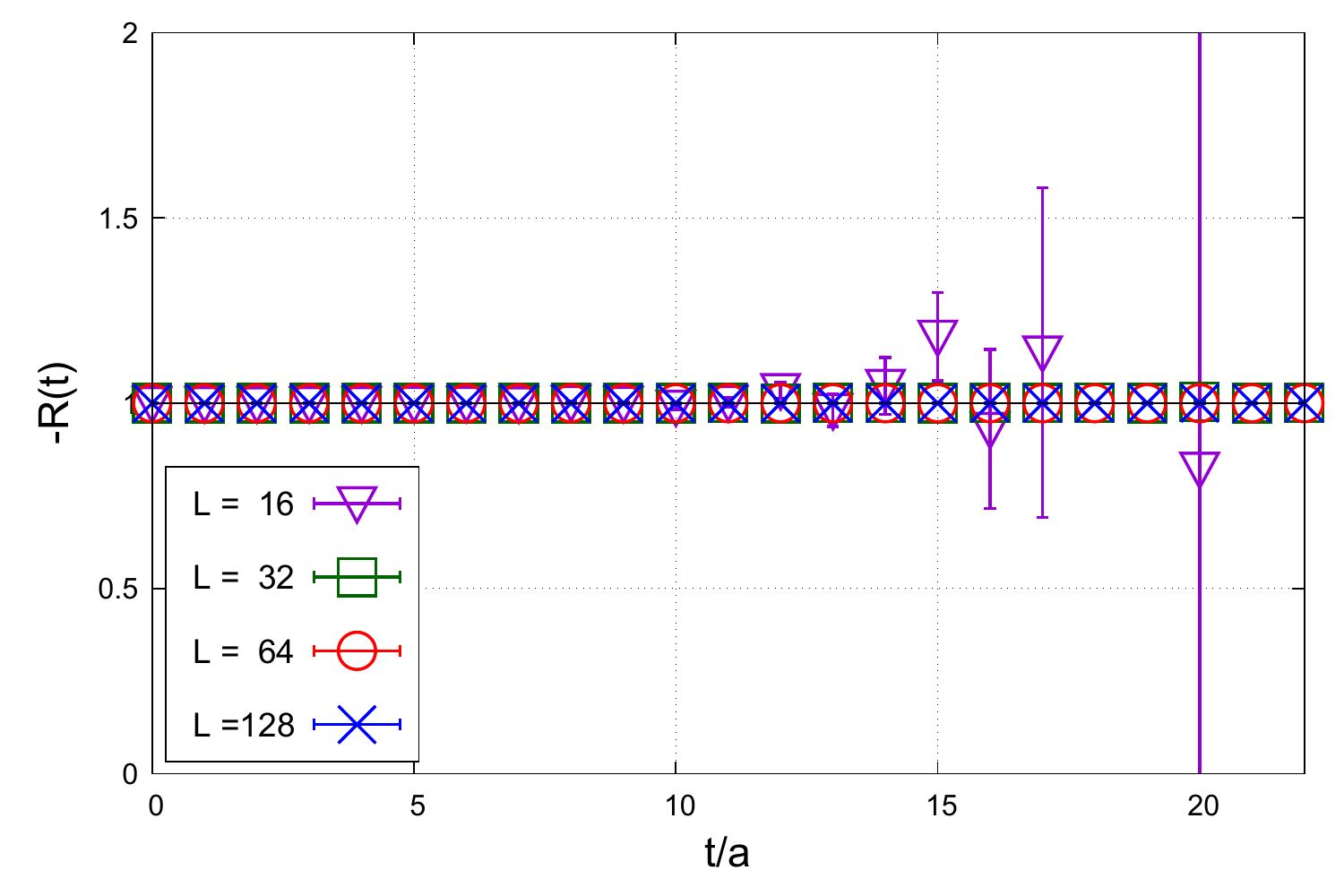}
 \caption{
Results of analyzing the mockup data for various $M_{\rm{pole}}^2$ and volume using our improved method.
The parameters in $G(q^2)$ are chosen as
$g_{1}=0.90/M_{\rm{pole}}^2,\hspace{2pt}g_{2}=-0.10/M_{\rm{pole}}^4$ for a given $M_{\rm pole}^2$.
Symbols are calculation results, and solid lines are exact values of $s_{1}=f_{1}+f_{0}g_{1}=-0.10/M_{\rm{pole}}^2$.
Left: $M_{\rm{pole}}^2$ dependence for $M_{\rm{pole}}^2 = 0.10, 0.20, 2.00$ with the fixed volume of $L=32$.
Right: Volume dependence for $L=16, 32, 64, 128$ with the fixed $M_{\rm pole}^2 = 0.10$.
  \label{fig:R(t)_Mpole_V_Imp}
 }
\end{figure}

We analyze the same mockup data in subsection~\ref{subsec:problem} using our improved method.
In this analysis, we choose the parameters in $G(q^2)$ for a given $M_{\rm pole}^2$
such that $g_{1}=0.90/M_{\rm{pole}}^2$ and $g_{2}=-0.10/M_{\rm{pole}}^4$, which satisfy $s_{2}=0$.
The left and right panels of Fig.~\ref{fig:R(t)_Mpole_V_Imp} show 
the $M_{\rm{pole}}^2$ and volume dependences of our improved method, respectively.
These results present that our method works well regardless of the volume as well as $M_{\rm{pole}}^2$.
Therefore, our improved method could be applicable to compute $f_{1}$ for small $M_{\rm{pole}}^2$ and volume.

\section{Application to actual lattice QCD data}
In this section, we will confirm that our improved model-independent method works well on actual lattice QCD data.

\subsection{Simulation parameters}\label{subsec:sim.para.}
We use 2+1 flavor gauge configurations generated by the PACS-CS Collaboration~\cite{Yamazaki:2012hi}
using the Iwasaki gauge action at $\beta=1.90$
and the nonperturbative $\order{a}$-improved Wilson quark action at $c_{\rm{SW}}=1.715$.
The ensemble parameters are shown in Table \ref{table:sim_param}.
The 3-point function $C_{\pi V\pi}(t,t_{\rm{sink}};r)$ is evaluated using Eq.~(\ref{eq:1pt3func}), and 
the $r^{2n}$ moment of the 3-point function is constructed using Eq.~(\ref{eq:1ptr^2n3func}).
The correlation functions are computed with the $Z(2)\otimes Z(2)$ random source~\cite{Boyle:2008yd}
and the value of $t_{\rm{sink}}$ is set to $22$.
The statistical error in the calculation is evaluated by the jackknife method.

\begin{table}[!h]
\begin{center}
\begin{tabular}{cccccccc}\hline\hline
$\beta$ & $L^3\times T$ & $L$[fm] & $a$[fm] & $M_\pi$[GeV] &
$N_{\rm conf}$ & $N_{\rm meas}$ \\
\hline
1.90 & 32$^3$$\times$48 & 2.9 & 0.090 & 0.51 & 80 & 192 \\
\hline\hline
\end{tabular}
\end{center}
\caption{
The bare coupling ($\beta$), lattice size ($L^3\times T$),
physical spatial extent ($L$[fm]), pion masses ($m_\pi$)
are tabulated.
$N_{\rm conf}$ and $N_{\rm meas}$ represent the number of the configurations
and the number of the measurement per configuration, respectively.
}
\label{table:sim_param}
\end{table}

\subsection{Results}
\begin{figure}[!th]
 \centering
 \includegraphics[width=91mm,pagebox=cropbox]{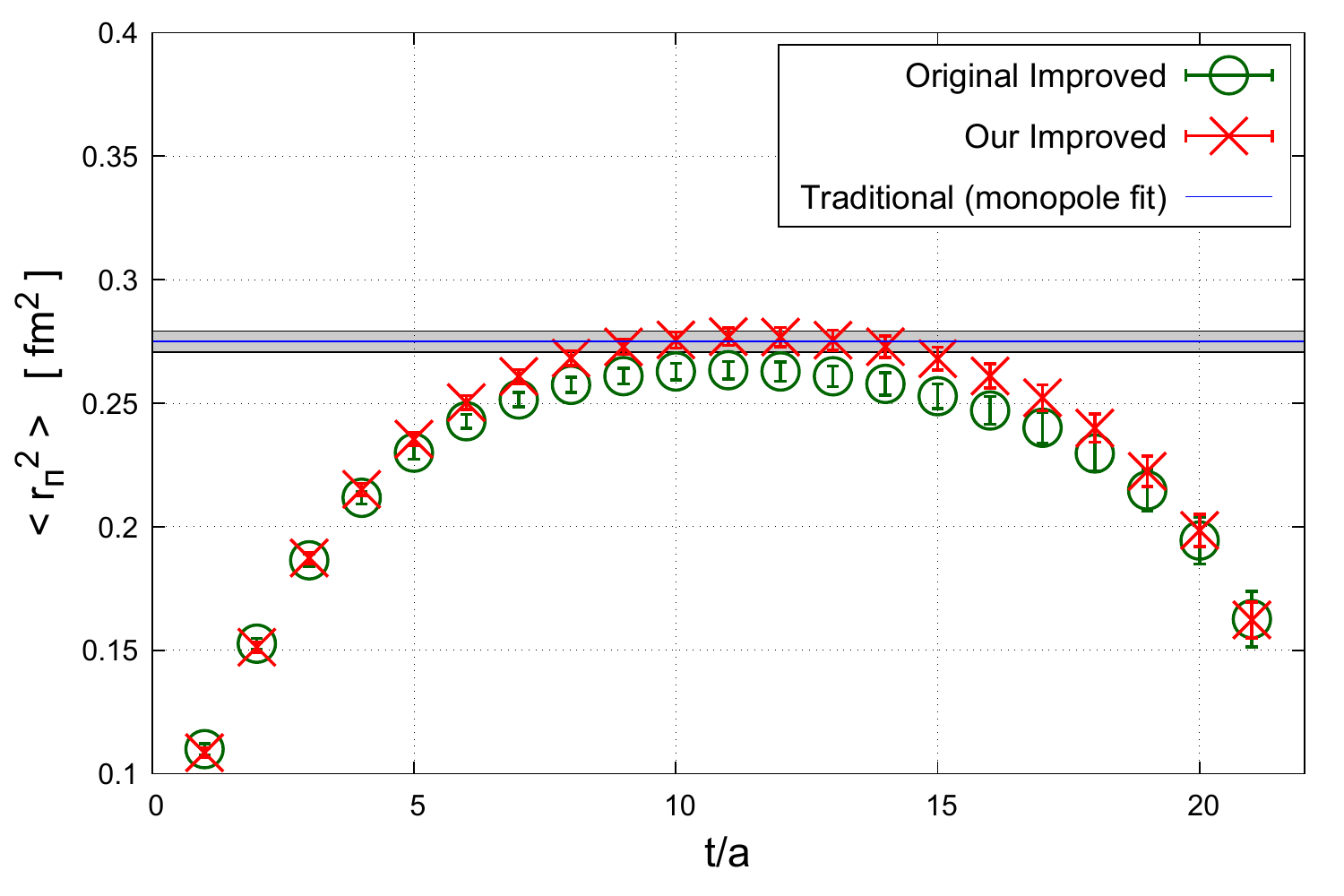}
 \caption{
Comparison of the pion charge radius in the actual lattice QCD
calculation using our method with the ones using the original model-independent method and also the traditional method.
The red symbols represent our model-independent method, and
the green symbols represent the original one.
Blue lines and gray band represent the central value and the statistical error using the traditional method with monopole fit.
  \label{fig:r_pi2_R(t)}
 }
\end{figure}

In this subsection, we present preliminary results with the original and our methods.
We choose $g_1 = 0$ as an example in our method.
We numerically search the value of $g_{2}$ satisfying $s_{2}=0$ by
changing $g_2$ with the fixed $g_1$.
Note that the value of $s_2$ can be obtained in a similar way to
Eq.~(\ref{eq:exp_Cn_new}) by replacing Eq.~(\ref{eq:defParaAlphaH}) with
$\alpha_{1}\tilde{\beta}_{0,1}+\alpha_{2}\tilde{\beta}_{0,2}+h=0$,
$\alpha_{1}\tilde{\beta}_{1,1}+\alpha_{2}\tilde{\beta}_{1,2}=0$, and
$\alpha_{1}\tilde{\beta}_{2,1}+\alpha_{2}\tilde{\beta}_{2,2}=1$.
From this search, $g_2 = -31.5$ is determined.

\par

Figure~\ref{fig:r_pi2_R(t)} presents the results of the pion charge radius $\expval{r_{\pi}^2}=-6f_{1}$
with the original and our methods calculated from $-6R(t)$ and $-6(R(t)-g_1)$, respectively, as a function of time.
In the original method (green circle symbols), 
the values of the pion charge radius in the middle time region are smaller than the result of the traditional method with the monopole fit (blue line and gray band).
The error of the monopole fit result is only statistical.
Although we did not estimate the systematic error of the traditional method yet,
the underestimate of the original method could be caused by the higher-order contamination discussed above.
It is a similar trend observed in our mockup data analysis presented in Fig.~\ref{fig:R(t)_Mpole_V_NotImp}.
On the other hand, the value of the pion charge radius becomes little larger and agrees with the traditional method within 1$\sigma$
for our method (red cross symbols) with $g_{1}=0, g_{2}=-31.5$.
The results of the analysis with other parameters $g_1$ and $g_2$ are shown in Table \ref{table:result}.
As the parameters, we choose $g_{1}=5.1$ to be close to the value of $1/M_{\rm{pole}}^2$ obtained from the traditional method,
and $g_{2} = -3.25$ is chosen to satisfy $s_{2}=0$ by the numerical search explained above.
The different choice of the parameters in our method gives a tiny effect in the charge radius.
The discrepancy between the two results with the different parameters is less than the statistical error as shown in the Table.
Furthermore, the result with $g_{1}=5.1$ and $g_{2}=-3.25$ is also consistent with the traditional method.
Thus, it shows the validity of our method on actual lattice QCD data.

\begin{table}[!h]
\caption{
Pion charge radius ($\expval{r_{\pi}^2}$[fm${}^{2}$])
obtained from the actual lattice QCD data with the model-independent methods at $t/a=12$.
The choice of $g_1=g_2=0$ corresponds to the original method,
and the others are determined from our method.
The result with the traditional method is also tabulated.
}
\begin{center}
\begin{tabular}{ccc}\hline\hline
$g_{1}$ & $g_{2}$ & $\expval{r_{\pi}^2}$ [fm${}^{2}$]  \\
\hline
0.0 &    0.0 & 0.2629(39) \\
0.0 & -31.5 & 0.2768(38) \\
5.1 & -3.25 & 0.2782(44) \\
\hline
    & traditional & 0.2750(42) \\
\hline\hline
\end{tabular}
\end{center}
\label{table:result}
\end{table}

\section{Summary}
We have discussed an improvement of the model-independent method to obtain the pion charge radius.
In the mockup data, 
we show that the original model-independent method has the large higher-order contamination
for small $M_{\rm{pole}}^2$ and volume.
The contamination is originated from a finite volume effect that is related to the convergence of the Taylor expansion of $F_{\pi}(q^2)$.
We propose a modified method to reduce the contamination by introducing an appropriate function
in the Taylor expansion.

\par

We have calculated the pion charge radius on actual lattice QCD data at the pion mass of 0.51 GeV
using the original and our methods, and compared these results with that using the traditional method.
We choose a quadratic function as the function $G(q^2)=1+g_{1}q^2+g_{2}q^4$,
where $g_{2}$ satisfying $s_{2}=0$ is determined by the numerical search with the fixed $g_{1}$.
The charge radius of the original model-independent method is underestimated
compared to the one of the traditional method with the monopole fit.
On the other hand,  our improved model-independent method is consistent with traditional method.
It is also found that the result with our method is stable against the different choice of the parameters.

\par

For a future work, we will estimate systematic errors in our improved method, such as the selection of $g_{1}$ and $g_{2}$,
although it is expected to be small from the preliminary results in the actual lattice QCD calculation.
We also need to estimate systematic error of the traditional method by using different fit forms other than the monopole form.
These estimations are necessary to confirm whether our improved method is beneficial.
Other future direction is to apply our method to more realistic calculations, {\it e.g.,} on larger volumes near the physical point
at smaller lattice spacings.

\section*{Acknowledgments}
Numerical calculations in this work were performed on Oakforest-PACS and
Wisteria/BDEC-01 (Odyssey) in Joint Center for Advanced High Performance Computing,
and on Cygnus in Center for Computational Sciences at University of Tsukuba
under Multidisciplinary Cooperative Research Program of Center for Computational
 Sciences, University of Tsukuba.
The calculation employed OpenQCD system\footnote{http://luscher.web.cern.ch/luscher/openQCD/}.
This work was supported in part by Grants-in-Aid
for Scientific Research from the Ministry of Education, Culture, Sports,
Science and Technology (No. 19H01892)
and JST, The Establishment of University Fellowships towards the creation of Science Technology Innovation, Grant Number JPMJFS2106.
This work was supported by the JLDG constructed over the SINET5 of NII.

\bibliography{reference}

\end{document}